\documentclass[aps,prl,showpacs,floatfix,twocolumn,superscriptaddress]{revtex4-1}

\pdfoutput=1

\usepackage{natbib} 
\usepackage{amsfonts} 
\usepackage{amsmath} 
\usepackage{amssymb} 
\usepackage{graphicx}
\usepackage{balance}

\begin{document}

\title{Diffusion-limited cluster aggregation: Impact of rotational diffusion}

\author{Swetlana Jungblut}
\email{swetlana.jungblut@tu-dresden.de}
\affiliation{Physikalische Chemie, TU Dresden, Bergstra{\ss}e 66b, 01069 Dresden, Germany}
\author{Jan-Ole Joswig}
\affiliation{Theoretische Chemie, TU Dresden, Bergstra{\ss}e 66c, 01069 Dresden, Germany}
\author{Alexander Eychm{\"u}ller}
\affiliation{Physikalische Chemie, TU Dresden, Bergstra{\ss}e 66b, 01069 Dresden, Germany}

\date{\today}

\begin{abstract}
Diffusion-limited cluster aggregation (DLCA) is a well established model for the formation of highly porous low-density non-equilibrium structures. One of the main conclusions of the previous studies considering this model is that the rotational diffusion of aggregating clusters does not change their structure characterized by a universal fractal dimension of $d_f=1.7-1.8$. In contradiction to this assumption, we demonstrate that the rotation movement of clusters significantly changes the structure of forming aggregates. The fractal dimension of rotating clusters is lower than the one found in the standard DLCA model and decreases with the increasing ratio of rotational and translational diffusion constants $D_r/D_t$, which offers a possibility to tune the structures of the aggregates below the conventional DLCA fractal dimension limit.

\end{abstract}


\maketitle

The model of diffusion-limited cluster aggregation (see, {\it e.g.}, Refs. \citenum{meakin:1992a, jullien:1992,lazzari:2016} for a review) was introduced more than 30 years ago \cite{kolb:1983,meakin:1983a}. Together with the closely related reaction-limited cluster aggregation model \cite{meakin:1987,meakin:1988}, DLCA provides a scenario for the process of non-equilibrium particle aggregation, which universality is widely accepted \cite{lin:1989,lin:1990a}. In this framework, aggregating nanoparticles form networks of variable density determined by the type of inter-particle interaction. The aggregates can be characterized by their fractal (Hausdorff) dimension, which measures how effectively they fill the available space. In the limiting case of non-interacting particles, which are currently assumed to form a structure with the lowest possible density, early Monte Carlo (MC) simulations of the aggregation process \cite{meakin:1984b} provided a value of $d_f=1.7-1.8$ for the fractal dimension of three-dimensional clusters. At the same time, a similar value was found experimentally by the analysis of two-dimensional images of the aggregates formed by gold colloidal particles \cite{weitz:1984a, weitz:1984b}. Later on, however, less dense fractal structures were observed by soot aggregation \cite{chakrabarty:2009,sanderchakrabarty:2010,chakrabartyReply:2010}. An explanation for the formation of aggregates with a fractal dimension below the DLCA limit is still under discussion. For instance, the anisotropic shape of the aggregates was suggested as a possible factor affecting the structure, but a recent study demonstrated that it does not influence the fractal dimension \cite{heinson:2010}. In addition, there are indications that the analysis of the two-dimensional images of experimentally obtained three-dimensional aggregates may overestimate their fractal dimension \cite{chakrabarty:2011a,martos:2017}. 

In this letter, we demonstrate that the aggregation of rotating clusters results in structures less dense than those obtained within the DLCA scheme. That is, we extend the conventional model of non-reversible DLCA to include the rotational diffusion of aggregates (rDLCA). The bulk of the previous studies concerning DLCA \cite{meakin:1985,hasmy:1995b,lattuada:2003a, lattuada:2003, rottereau:2004a, rottereau:2004b,diezOrrite:2005,babu:2008,heinson:2010} modeled Brownian dynamics of the aggregates in MC simulations and, in doing so, omitted the orientational diffusion of the clusters. The reason for this neglecting may be traced back to one of the earlier studies \cite{meakin:1988}, which equated rotations with selecting a random relative orientation of the aggregating clusters, compared the results with those of purely translationally diffusing clusters, and concluded that the rotational effects on DLCA are negligible. 
Hence, the following experimental investigations \cite{lindsay:1988,lindsay:1989} considered rotational diffusion only as a factor influencing the measurements of the dynamic scattering coefficient but not the structure of the aggregates. 
In contrast to previous studies, we find that an explicit implementation of the aggregates' rotational diffusion yields clusters, which are less compact and more anisotropic in shape than those produced by the translational diffusion only.

We studied the evolution of the system in a canonical $NVT$ ensemble with temperature controlled by the Langevin thermostat for rigid body dynamics \cite{davidchack:2015}, which couples on the translational as well as rotational degrees of freedom. Throughout the paper, all distances are given in units of sphere diameter $\sigma$ and the time in $\tau = \sigma \sqrt{m/k_{\rm B}T}$ with $k_{\rm B}$ being the Boltzmann constant and $T$ the temperature. The thermal energy of the system $k_{\rm B}T$ and the mass of a single particle $m$ are set to unity. The mass of an aggregate is equal to $n$, the number of particles it contains. The friction coefficients of the Langevin thermostats are related to the single-particle translational and rotational diffusion constants via $\gamma_{t|r}=k_{\rm B}T/D_{t|r}$. The evolution of the system is integrated with a time step of $\Delta t = 0.001$ and the diffusion constants are set to  $D_t = 0.1$ and $D_r=0.5D_t$. 
The particles were confined to a cubic box with the edge $L=60$ and periodic boundary conditions were applied in all directions. The number of particles $N$ varied between $640$ and $3600$, yielding a set of volume fractions $\phi_i$=$\{0.0015514$, $0.00193925$, $0.00232711$, $0.00290888$, $0.00387851$, $0.00484814$, $0.00581776$, $0.00678739$, $0.00775702$, $0.00872665\}$.    
Simulating non-reversible aggregation, we assume that particles and clusters collide inelastically at the given cutoff distance, which was set to $r_c=1$, and continue their movement as rigid bodies with the translational and angular momenta conserved. The rotations of the aggregates were implemented using quaternions \cite{miller:2002}.
For each density, we simulate the aggregation of $100$ realizations of initially disordered systems obtained from equilibrated simulations of strongly repulsive particles. Initial particle velocities were chosen from the Maxwell-Boltzmann distribution. Each run continues until all particles connect into a single aggregate. Along the runs, we monitor the size and the shape of the clusters containing more than one particle.   

\begin{figure}[tb]
\begin{center}
\includegraphics[clip=,width=0.99\columnwidth]{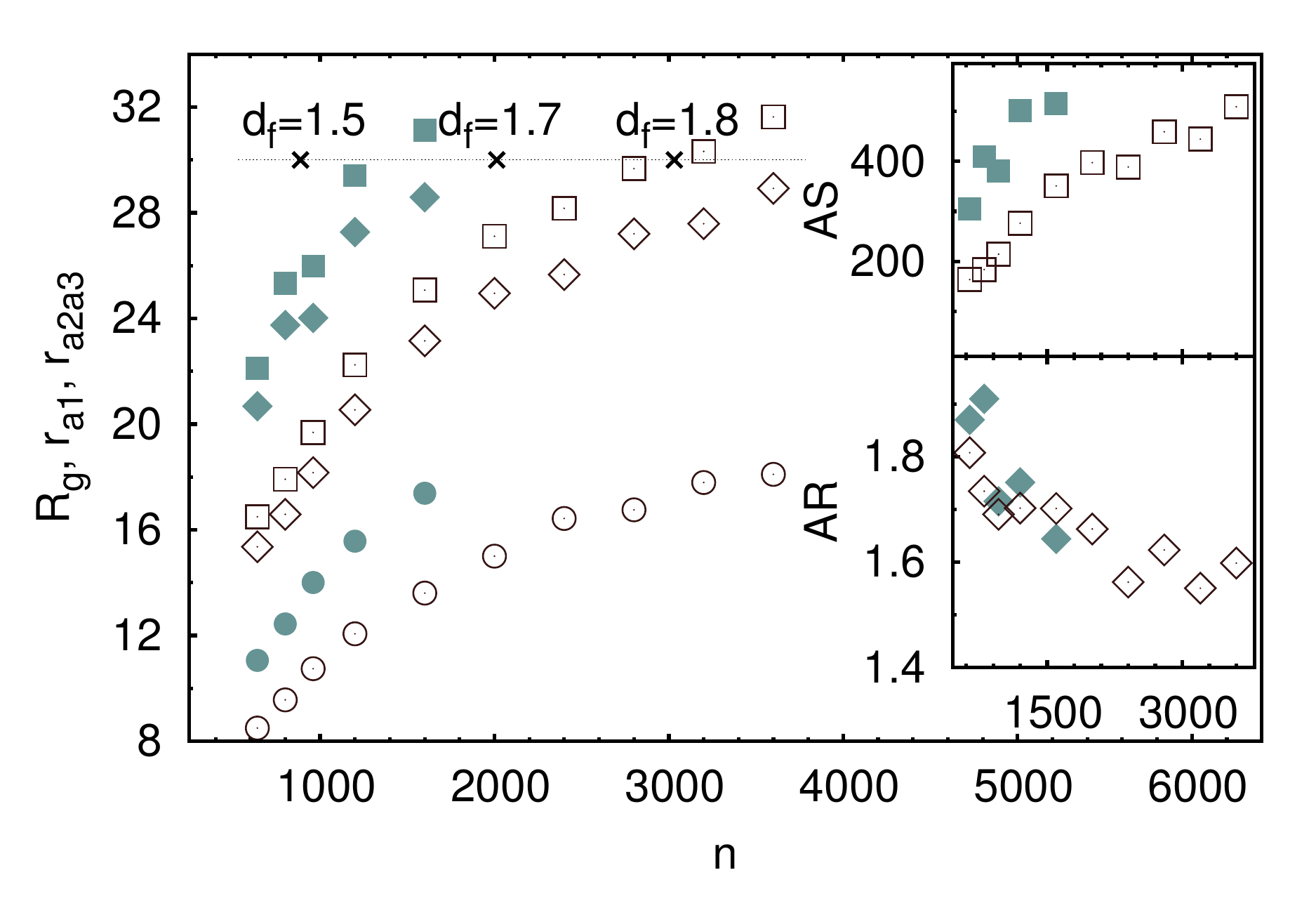} 
\caption{\label{rofgAsymm} (Color online) Radius of gyration as a function of the number of particles in the aggregate with (solid red symbols) and without (open blue symbols) rotations. We differentiate between the radius of gyration of the whole cluster, $R_g$ (squares), the radius of gyration in the plane perpendicular to the main principal axis (circles), and the average over the radii around the other two axes (diamonds). The horizontal line demonstrates the relation of the aggregate size to the dimensions of the simulation box, $L/2$. The points on this line, computed with Eq.~\ref{ncofdf}, stand for the expected onset of percolation for clusters of various fractal dimensionality (as labeled). Insets: Aspherity parameters (top) and aspect ratios (bottom) of the aggregates. Note that the aspherity parameter is zero for a spherical object or a spherically symmetric distribution of particles, while the aspect ratio of a spherical shape is one. } 
\end{center}
\end{figure}
The first differences between the structures of the aggregates assembled with and without rotational motion, which we observe in our analysis, are related to their overall dimensions. In Fig.~\ref{rofgAsymm}, we compare the values of the radii of gyration of the aggregates, 
\begin{equation}
R_g^{2} = \lambda_1^2 + \lambda_2^2 + \lambda_3^2,
\end{equation} 
computed from the eigenvalues ($\lambda_1\geq \lambda_2 \geq\lambda_3$) of the gyration tensor, which is constructed from the positions of all particles of an aggregate. 
The eigenvalues of the gyration tensor can be additionally used to determine the aspherity of the clusters \cite{theodorou:1985},  
\begin{equation}
AS = \lambda_1^2-0.5(\lambda_2^2 + \lambda_3^2), 
\end{equation}
which approaches zero for a perfectly spherical object or a spherically symmetric distribution of particles in a cluster.
Furthermore, the gyration tensor can be mapped on the inertia tensor \cite{vymetal:2011} to compute the radii of gyration around the principal axes of an aggregate. In the following, we consider the radius around the axis corresponding to the largest eigenvalue of the inertia tensor, $a1$, separately and average over the radii around the other two axes, $a2$ and $a3$,  
\begin{eqnarray}
&&r_{a1}^2 = \lambda_2^2 + \lambda_3^2 \nonumber \\
&&r_{a2a3}^2 = 0.5(r_{a2}^2+r_{a3}^2) = \lambda_1^2 + 0.5(\lambda_2^2 + \lambda_3^2).\label{radii}
\end{eqnarray}
The aspect ratio of the aggregating clusters is given by the ratio of these two radii: 
\begin{equation}
AR = \sqrt{\frac{\lambda_1^2 + 0.5(\lambda_2^2 + \lambda_3^2)}{\lambda_2^2 + \lambda_3^2}}. 
\end{equation}
Combining the values of the aspect ratio of final aggregates with their aspherity parameters, we conclude that the distribution of particles inside the clusters becomes less spherically symmetric on growing, while the elongation of the clusters decreases. 
Figure~\ref{rofgAsymm} demonstrates that the effects are more pronounced for rDLCA clusters than for the conventional DLCA aggregates. Such behavior is characteristic for a fractal system approaching percolation transition, at which the aggregates start to fill the space homogeneously.    
Restricted to the simulation box, the transition occurs when the clusters become connected to themselves through periodic boundary conditions, {\it i.~e.}, when their radius of gyration becomes equal to $L/2$. Note that, in our simulations, we do not yield such connections directly but infer their occurrence from averaging over different realizations of the aggregates. 
Using the relation between the initial volume fraction and the radius of gyration of aggregates at percolation, $R_g^c=(\sigma/2)\phi^{1/(d_f-3)}$, derived in previous investigations of DLCA \cite{carpineti:1992, bibette:1992a}, we can estimate the number of particles needed to be initially present in a system of given size to observe percolation transition, 
\begin{equation}
n^c=\frac{6}{\pi}L^{d_f}. \label{ncofdf}
\end{equation}
Consequently, we compare the predictions for different values of fractal dimension with the results obtained in simulations. 
Figure~\ref{radii} shows that the dimensions of rDLCA clusters become comparable with the simulation box size at initial densities corresponding to a fractal dimension that is lower than the one of DLCA aggregates. 

In principle, we do not have to deduce the fractal dimension of the aggregates from their behavior at the percolation threshold, particularly since it is well known that the fractal dimension of the clusters increases as they start to form homogeneous networks \cite{rottereau:2004a, rottereau:2004b}. The conventional route for the estimation of the fractal dimension of the aggregates utilizes the relation between their size and radius of gyration,    
\begin{equation}
R_g=k n^{1/d_f},\label{rofgscaling}
\end{equation}
where $k$ is a factor connected to the mass density and is used along with $d_f$ to fit the relation to the data. 
\begin{figure}[tb]
\begin{center}
\includegraphics[clip=,width=0.99\columnwidth]{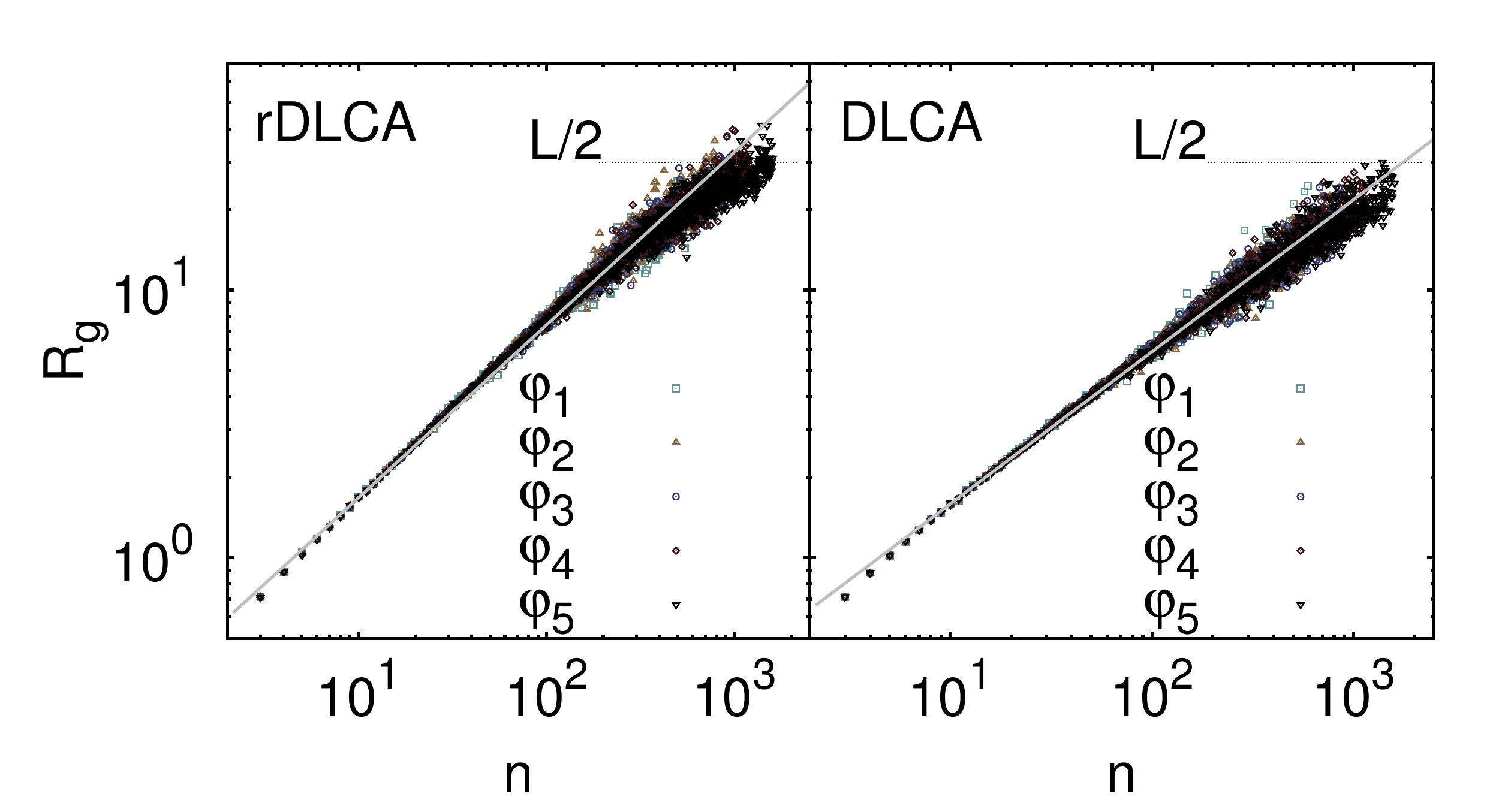} 
\caption{\label{rofg_fit} (Color online) Radius of gyration with (left) and without (right) rotations as a function of the aggregate size. Up to a certain length scale dependent on the initial particle volume fraction the radius of gyration is independent on the density and scales with the cluster size according to Eq.~(\ref{rofgscaling}), presented by gray lines with $\{d_f=1.55$, $k=0.38\}$ and $\{d_f=1.76$, $k=0.43\}$ as representative fits for rDLCA and DLCA clusters, respectively. All values of the fractal dimension are given in Table~\ref{table}.} 
\end{center}
\end{figure}
In order to get a sufficient amount of aggregate sizes, we monitored the shapes of all clusters found in the system along the time spent at the formation of the final aggregate. In Fig.~\ref{rofg_fit}, we plot the raduis of gyration of clusters as a function of their size for a number of initial densities. Evidently, there is a range of cluster sizes in which the dimensions of the aggregates are independent on the volume fraction of particles initially present in the system. The connection to Eq.~(\ref{rofgscaling}) is illustrated by the lines with representative fitted parameters $d_f$ and $k$. Table~\ref{table} summarizes all fractal dimensions obtained from fitting Eq.~(\ref{rofgscaling}) either to all data points or to the data in the range $5<n<600$. We attribute the slight increase of the fractal dimension with the particle volume fraction, obtained by the fit to all data points, to the onset of percolation, at which the fractal dimension of clusters changes its value to $d_f^{\rm perc}=2.5$. If, however, we stay in the size range far away from the transition by restricting the fitting range to $5<n<600$, the fractal dimension becomes nearly constant for all initial volume fractions. The value of the fractal dimension yielded from the fits conforms with the standard DLCA value, $d_f=1.7-1.8$, only if the clusters are not allowed to rotate. Otherwise, in agreement with the observation of final aggregates reaching percolation transition, the fractal dimension has a smaller value of $d_f=1.5-1.6$. Furthermore, an increase of the rotational diffusion, as exemplified for one value of the initial density, yields further decrease of the fractal dimension. We also use the data to analyze the impact of the projection of three-dimensional structures on two-dimensional images which is a standard practice to determine the fractal dimension experimentally. In doing so, we project the aggregates on the planes perpendicular to their principal axes. The radii of gyration of such projections represent the limiting cases in which the aggregates are aligned either parallel or perpendicular to the image surface. Otherwise, an average over all possible orientations yields the standard three-dimensional $R_g$. Table~\ref{table} demonstrates that the largest deviation in the estimated values of fractal dimension occurs for the axis associated with the largest principal moment of inertia. This variation is, however, less pronounced than the effects due to rotations. 
\begin{table}[tb]
\caption{
  \label{table} 
  Fractal dimensions obtained from fitting Eq.~(\ref{rofgscaling}) to the data points with $n > 5$ ($d_f^{\rm all}$) and restricted to $5<n<600$ ($d_f$). Last two columns ($5<n<600$) provide the fractal dimensions estimated from the projection of the aggregates on the planes normal to their main principal axis ($d_f^{a1}$) and averaged over the other two axes ($d_f^{a2a3}$). Uncertainty of all fits is of the order $\pm 0.001$.}
\begin{ruledtabular}
  \begin{tabular*}{0.5\textwidth}{@{\extracolsep{\fill}}lrrrr}
    & $d_f^{\rm all} $ & $d_f $& $d_f^{a1}$& $d_f^{a2a3}$ \\
    \hline
    $D_r=2D_t$ & & & &  \\  
    $\phi_1$  & $1.485 $& $1.465 $ & $1.535 $& $1.462 $\\
    \hline
    $D_r=D_t$ & & & &  \\   
    $\phi_1$  & $1.513 $& $1.498 $ & $1.532 $& $1.498 $\\
    \hline
     $D_r=0.5D_t$& & & & \\
    $\phi_1$  & $1.577 $& $1.565 $& $1.553 $& $1.570 $\\
    $\phi_2$  & $1.570 $ & $1.548 $& $1.562 $& $1.551 $\\
    $\phi_3$  & $1.589 $ & $1.547 $& $1.571 $& $1.548 $\\
    $\phi_4$  & $1.603 $ & $1.557 $& $1.557 $& $1.562 $\\
    $\phi_5$  & $1.643 $ & $1.575 $& $1.556 $& $1.582 $\\

  \hline
    $D_r=0$ & & & &  \\
    $\phi_1$  & $1.764 $ & $1.757 $& $1.715 $& $1.768 $\\
    $\phi_2$  & $1.781 $ & $1.771 $& $1.712 $& $1.784 $\\
    $\phi_3$  & $1.774 $ & $1.757 $& $1.728 $& $1.766 $\\
    $\phi_4$  & $1.781 $ & $1.764 $& $1.718 $& $1.776 $\\
    $\phi_5$  & $1.815 $ & $1.790 $& $1.738 $& $1.804 $\\
    $\phi_6$  & $1.825 $ & $1.794 $& $1.741 $& $1.807 $\\
    $\phi_7$  & $1.846 $ & $1.800 $& $1.745 $& $1.815 $\\
    $\phi_8$  & $1.867 $ & $1.812 $& $1.749 $& $1.828 $\\
    $\phi_9$  & $1.881 $ & $1.817 $& $1.749 $& $1.834 $\\
    $\phi_{10}$  & $1.896 $ & $1.831 $& $1.765 $& $1.849 $\\
  \end{tabular*}
\end{ruledtabular}
\end{table}

\begin{figure}[tb]
\begin{center}
\includegraphics[clip=,width=0.99\columnwidth]{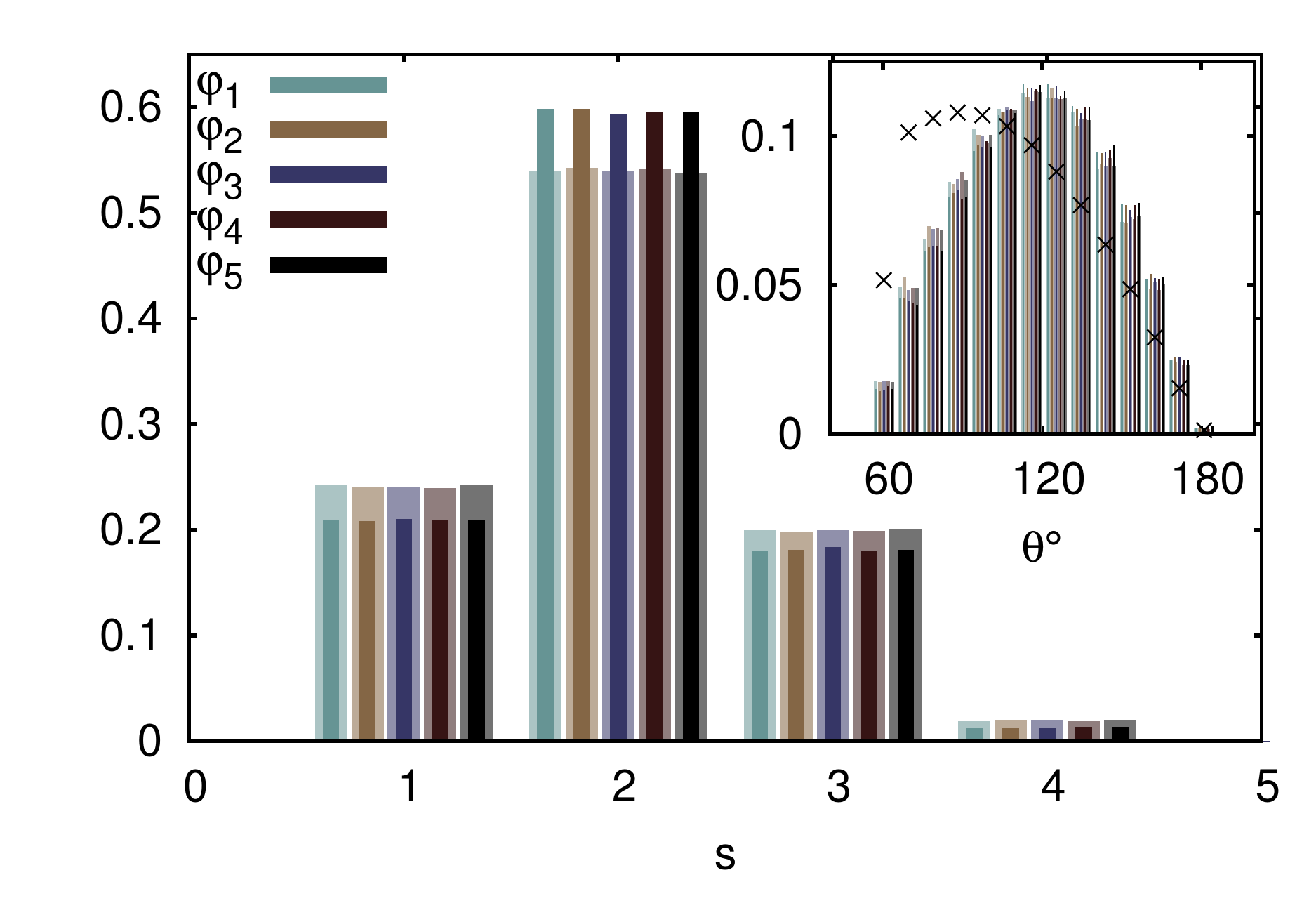} 
\caption{\label{neighbors} (Color online) Distribution of the number of neighbors per particle, $s$, in final aggregates obtained with (solid impulses) and without (transparent boxes) rotations. Inset: Distribution of the angles between the vectors connecting a particle (with $s=2$) to its neighbors. Additional points stand for the random distribution truncated at $\theta=60^{\circ}$. } 
\end{center}
\end{figure}
Next, we probe the structure of the final aggregates. The local arrangement of particles is reflected in the number of neighbors of a particle. Figure \ref{neighbors} demonstrates that, as expected, the particles mainly arrange themselves into interlacing chains with a few loose ends ($s=1$) and junctions ($s=3$). This local structure does not significantly depend on the initial density of the system. The removal of rotations, however, leads to an increase of the number of loose ends and junctions, which corresponds to the aggregates with a higher fractal dimension. 
We further selected particles forming chains ($s=2$) and sampled the angles between the vectors connecting a particle to its neighbors. The resulting distribution, also presented in Fig.~\ref{neighbors}, is shifted to larger angles in comparison to the distribution of randomly distributed contacts (restricted to angles producing no overlap between the neighbors of a particle). The shift is slightly more pronounced in chains formed by rDLCA and indicates that the chains are more linear than one would expect if the contact were distributed randomly.      
   
\begin{figure}[tb]
\begin{center}
\includegraphics[clip=,width=0.99\columnwidth]{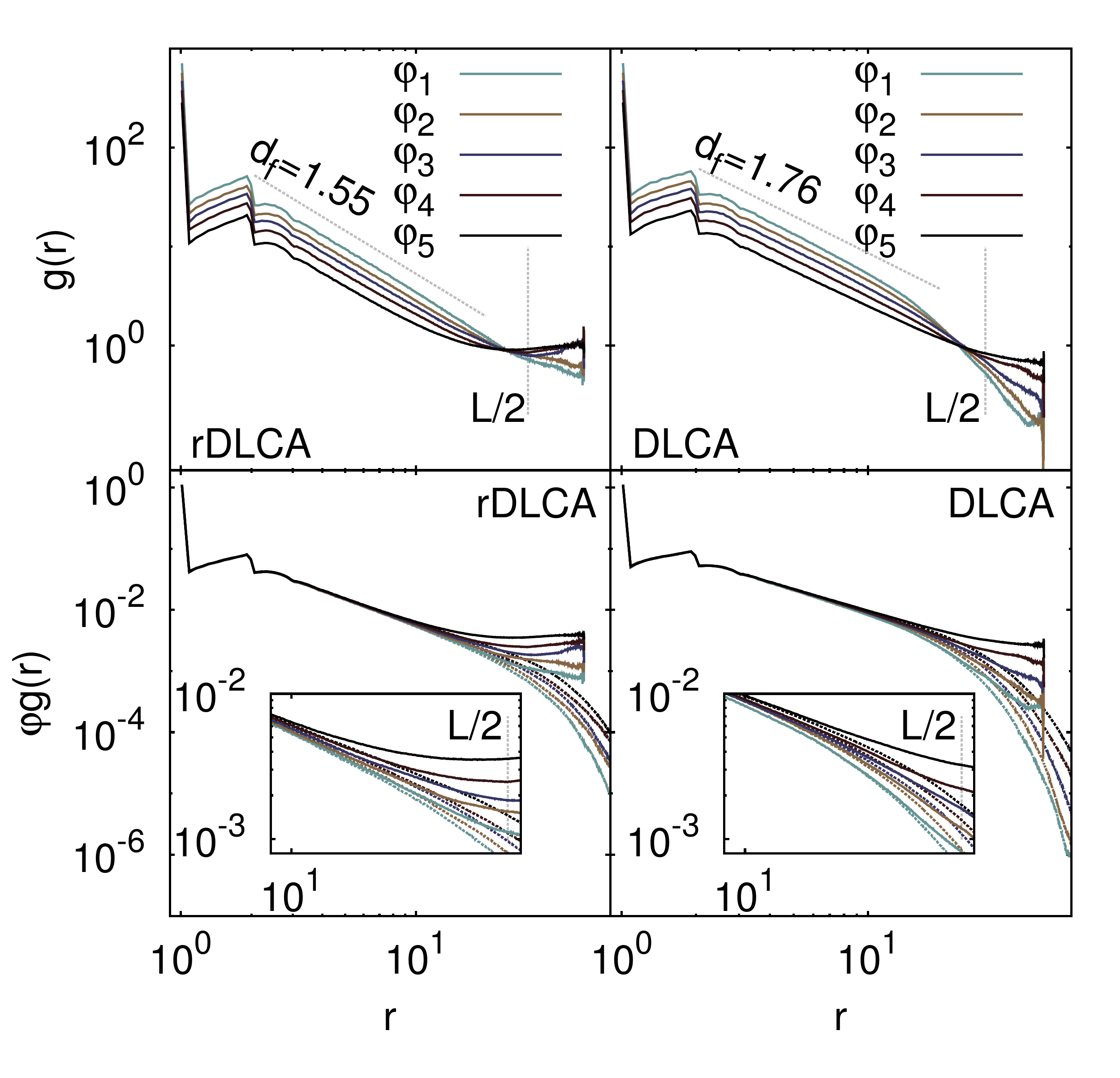} 
\caption{\label{gofr} (Color online) Top: Radial distribution function $g(r)$ computed in simulations with (left frame) and without (right frame) rotational diffusion. Vertical dotted line indicates the dimension of the simulation box, $L/2$. Straight dotted lines demonstrate linear scaling $g(r)\propto r^{3-d_f}$ on the intermediate length scales. Bottom: Pair distribution function $\phi g(r)$ computed with (solid lines) and without (broken lines) application of periodic boundary conditions. Insets: Length scale for the onset of deviations between the function values computed with and without periodic boundary conditions decreases with increasing initial volume fraction. } 
\end{center}
\end{figure}
The overall structure of the aggregates is reflected in the radial distribution function $g(r)$. Data presented in Fig.~\ref{gofr} confirms our observations on the difference of structures obtained with and without rotational diffusion at all densities considered. Locally, it is more probable to find another particle in the vicinity of a given particle for non-rotating aggregates, which is a sign of a locally denser system with a higher fractal dimension. In contrast, at the larger scale, it is more probable to find particles further away from each other if the aggregates are allowed to rotate. Hence, in this case, the aggregates are less compact and require more space. The large scale behavior indicates to the onset of percolation at larger volume fractions, where the radial distribution function saturates to unity at $L/2$. In order to emphasize the independence of the local structure of aggregates on the initial volume fraction, we plot, in the bottom frames of Fig.~\ref{gofr}, the pair distribution function $\phi g(r)$. Evidently, up to a certain distance dependent on the initial density, the structures are identical. On the larger scale, however, the probability to find a particle at a given distance increases with the initial volume fraction. By considering pair distribution functions computed with and without periodic boundary conditions, we attribute this increase to both the increasing dimensions of the aggregates and their overlap through periodic boundary conditions. These observations further explain the spurious dependence of the fractal dimension on the initial volume fraction seen in Table~\ref{table} when all aggregates are considered in the fit and its disappearance when only clusters in the size range, in which the pair distribution function is independent on the initial density, are taken into account. 

In summary, we have studied irreversible aggregation of particles into disordered structures in a well established DLCA model and demonstrated that taking into account the rotational diffusion of aggregating clusters decreases their fractal dimension below the standard DLCA limit. The results are intuitive but up to now the rotation was not considered to have a significant impact on the process of particle aggregation. Without rotations, we recover the classical value of fractal dimension, widely assumed as the limit of lowest density that can be achieved for fractal aggregates without introducing inter-particle attraction \cite{kim:2000a,puertas:2001,kim:2003}. We have shown, however, that an increase of the magnitude of the rotational in relation to the translational diffusion further decreases the fractal dimension of the aggregates. 

\begin{acknowledgments}
We gratefully acknowledge financial support from the European Research Council (ERC-2013-AdG AEROCAT). The computations were performed on an HPC system at the Center for Information Services and High Performance Computing (ZIH) at TU Dresden within the project QDSIM.
\end{acknowledgments}

%


\begin{thebibliography}{37}%
\makeatletter
\providecommand \@ifxundefined [1]{%
 \@ifx{#1\undefined}
}%
\providecommand \@ifnum [1]{%
 \ifnum #1\expandafter \@firstoftwo
 \else \expandafter \@secondoftwo
 \fi
}%
\providecommand \@ifx [1]{%
 \ifx #1\expandafter \@firstoftwo
 \else \expandafter \@secondoftwo
 \fi
}%
\providecommand \natexlab [1]{#1}%
\providecommand \enquote  [1]{``#1''}%
\providecommand \bibnamefont  [1]{#1}%
\providecommand \bibfnamefont [1]{#1}%
\providecommand \citenamefont [1]{#1}%
\providecommand \href@noop [0]{\@secondoftwo}%
\providecommand \href [0]{\begingroup \@sanitize@url \@href}%
\providecommand \@href[1]{\@@startlink{#1}\@@href}%
\providecommand \@@href[1]{\endgroup#1\@@endlink}%
\providecommand \@sanitize@url [0]{\catcode `\\12\catcode `\$12\catcode
  `\&12\catcode `\#12\catcode `\^12\catcode `\_12\catcode `\%12\relax}%
\providecommand \@@startlink[1]{}%
\providecommand \@@endlink[0]{}%
\providecommand \url  [0]{\begingroup\@sanitize@url \@url }%
\providecommand \@url [1]{\endgroup\@href {#1}{\urlprefix }}%
\providecommand \urlprefix  [0]{URL }%
\providecommand \Eprint [0]{\href }%
\providecommand \doibase [0]{http://dx.doi.org/}%
\providecommand \selectlanguage [0]{\@gobble}%
\providecommand \bibinfo  [0]{\@secondoftwo}%
\providecommand \bibfield  [0]{\@secondoftwo}%
\providecommand \translation [1]{[#1]}%
\providecommand \BibitemOpen [0]{}%
\providecommand \bibitemStop [0]{}%
\providecommand \bibitemNoStop [0]{.\EOS\space}%
\providecommand \EOS [0]{\spacefactor3000\relax}%
\providecommand \BibitemShut  [1]{\csname bibitem#1\endcsname}%
\let\auto@bib@innerbib\@empty
\bibitem [{\citenamefont {Meakin}(1992)}]{meakin:1992a}%
  \BibitemOpen
  \bibfield  {author} {\bibinfo {author} {\bibfnamefont {P.}~\bibnamefont
  {Meakin}},\ }\href@noop {} {\bibfield  {journal} {\bibinfo  {journal}
  {CCACAA}\ }\textbf {\bibinfo {volume} {65}},\ \bibinfo {pages} {237}
  (\bibinfo {year} {1992})}\BibitemShut {NoStop}%
\bibitem [{\citenamefont {Jullien}(1992)}]{jullien:1992}%
  \BibitemOpen
  \bibfield  {author} {\bibinfo {author} {\bibfnamefont {R.}~\bibnamefont
  {Jullien}},\ }\href@noop {} {\bibfield  {journal} {\bibinfo  {journal}
  {CCACAA}\ }\textbf {\bibinfo {volume} {65}},\ \bibinfo {pages} {215}
  (\bibinfo {year} {1992})}\BibitemShut {NoStop}%
\bibitem [{\citenamefont {Lazzari}\ \emph {et~al.}(2016)\citenamefont
  {Lazzari}, \citenamefont {Nicoud}, \citenamefont {Jaquet}, \citenamefont
  {Lattuada},\ and\ \citenamefont {Morbidelli}}]{lazzari:2016}%
  \BibitemOpen
  \bibfield  {author} {\bibinfo {author} {\bibfnamefont {S.}~\bibnamefont
  {Lazzari}}, \bibinfo {author} {\bibfnamefont {L.}~\bibnamefont {Nicoud}},
  \bibinfo {author} {\bibfnamefont {B.}~\bibnamefont {Jaquet}}, \bibinfo
  {author} {\bibfnamefont {M.}~\bibnamefont {Lattuada}}, \ and\ \bibinfo
  {author} {\bibfnamefont {M.}~\bibnamefont {Morbidelli}},\ }\href@noop {}
  {\bibfield  {journal} {\bibinfo  {journal} {Adv. Coll. Inter. Sci.}\ }\textbf
  {\bibinfo {volume} {235}},\ \bibinfo {pages} {1} (\bibinfo {year}
  {2016})}\BibitemShut {NoStop}%
\bibitem [{\citenamefont {Kolb}\ \emph {et~al.}(1983)\citenamefont {Kolb},
  \citenamefont {Botet},\ and\ \citenamefont {Jullien}}]{kolb:1983}%
  \BibitemOpen
  \bibfield  {author} {\bibinfo {author} {\bibfnamefont {M.}~\bibnamefont
  {Kolb}}, \bibinfo {author} {\bibfnamefont {B.}~\bibnamefont {Botet}}, \ and\
  \bibinfo {author} {\bibfnamefont {R.}~\bibnamefont {Jullien}},\ }\href@noop
  {} {\bibfield  {journal} {\bibinfo  {journal} {Phys. Rev. Lett.}\ }\textbf
  {\bibinfo {volume} {51}},\ \bibinfo {pages} {1123} (\bibinfo {year}
  {1983})}\BibitemShut {NoStop}%
\bibitem [{\citenamefont {Meakin}(1983)}]{meakin:1983a}%
  \BibitemOpen
  \bibfield  {author} {\bibinfo {author} {\bibfnamefont {P.}~\bibnamefont
  {Meakin}},\ }\href@noop {} {\bibfield  {journal} {\bibinfo  {journal} {Phys.
  Rev. Lett.}\ }\textbf {\bibinfo {volume} {51}},\ \bibinfo {pages} {1119}
  (\bibinfo {year} {1983})}\BibitemShut {NoStop}%
\bibitem [{\citenamefont {Meakin}\ and\ \citenamefont
  {Family}(1987)}]{meakin:1987}%
  \BibitemOpen
  \bibfield  {author} {\bibinfo {author} {\bibfnamefont {P.}~\bibnamefont
  {Meakin}}\ and\ \bibinfo {author} {\bibfnamefont {F.}~\bibnamefont
  {Family}},\ }\href@noop {} {\bibfield  {journal} {\bibinfo  {journal} {Phys.
  Rev. A}\ }\textbf {\bibinfo {volume} {36}},\ \bibinfo {pages} {5498}
  (\bibinfo {year} {1987})}\BibitemShut {NoStop}%
\bibitem [{\citenamefont {Meakin}\ and\ \citenamefont
  {Family}(1988)}]{meakin:1988}%
  \BibitemOpen
  \bibfield  {author} {\bibinfo {author} {\bibfnamefont {P.}~\bibnamefont
  {Meakin}}\ and\ \bibinfo {author} {\bibfnamefont {F.}~\bibnamefont
  {Family}},\ }\href@noop {} {\bibfield  {journal} {\bibinfo  {journal} {Phys.
  Rev. A}\ }\textbf {\bibinfo {volume} {38}},\ \bibinfo {pages} {2110}
  (\bibinfo {year} {1988})}\BibitemShut {NoStop}%
\bibitem [{\citenamefont {Lin}\ \emph {et~al.}(1989)\citenamefont {Lin},
  \citenamefont {Lindsay}, \citenamefont {Weitz}, \citenamefont {Ball},
  \citenamefont {Klein},\ and\ \citenamefont {Meakin}}]{lin:1989}%
  \BibitemOpen
  \bibfield  {author} {\bibinfo {author} {\bibfnamefont {M.~Y.}\ \bibnamefont
  {Lin}}, \bibinfo {author} {\bibfnamefont {H.~M.}\ \bibnamefont {Lindsay}},
  \bibinfo {author} {\bibfnamefont {D.~A.}\ \bibnamefont {Weitz}}, \bibinfo
  {author} {\bibfnamefont {R.~C.}\ \bibnamefont {Ball}}, \bibinfo {author}
  {\bibfnamefont {R.}~\bibnamefont {Klein}}, \ and\ \bibinfo {author}
  {\bibfnamefont {P.}~\bibnamefont {Meakin}},\ }\href@noop {} {\bibfield
  {journal} {\bibinfo  {journal} {Nature}\ }\textbf {\bibinfo {volume} {339}},\
  \bibinfo {pages} {360} (\bibinfo {year} {1989})}\BibitemShut {NoStop}%
\bibitem [{\citenamefont {Lin}\ \emph {et~al.}(1990)\citenamefont {Lin},
  \citenamefont {Lindsay}, \citenamefont {Weitz}, \citenamefont {Klein},
  \citenamefont {Ball},\ and\ \citenamefont {Meakin}}]{lin:1990a}%
  \BibitemOpen
  \bibfield  {author} {\bibinfo {author} {\bibfnamefont {M.~Y.}\ \bibnamefont
  {Lin}}, \bibinfo {author} {\bibfnamefont {H.~M.}\ \bibnamefont {Lindsay}},
  \bibinfo {author} {\bibfnamefont {D.~A.}\ \bibnamefont {Weitz}}, \bibinfo
  {author} {\bibfnamefont {R.}~\bibnamefont {Klein}}, \bibinfo {author}
  {\bibfnamefont {R.~C.}\ \bibnamefont {Ball}}, \ and\ \bibinfo {author}
  {\bibfnamefont {P.}~\bibnamefont {Meakin}},\ }\href@noop {} {\bibfield
  {journal} {\bibinfo  {journal} {J. Phys.: Condens. Matter}\ }\textbf
  {\bibinfo {volume} {2}},\ \bibinfo {pages} {3093} (\bibinfo {year}
  {1990})}\BibitemShut {NoStop}%
\bibitem [{\citenamefont {Meakin}\ \emph {et~al.}(1984)\citenamefont {Meakin},
  \citenamefont {Majid}, \citenamefont {Havlin},\ and\ \citenamefont
  {Stanley}}]{meakin:1984b}%
  \BibitemOpen
  \bibfield  {author} {\bibinfo {author} {\bibfnamefont {P.}~\bibnamefont
  {Meakin}}, \bibinfo {author} {\bibfnamefont {I.}~\bibnamefont {Majid}},
  \bibinfo {author} {\bibfnamefont {S.}~\bibnamefont {Havlin}}, \ and\ \bibinfo
  {author} {\bibfnamefont {H.~E.}\ \bibnamefont {Stanley}},\ }\href@noop {}
  {\bibfield  {journal} {\bibinfo  {journal} {J. Phys. A: Math. Gen.}\ }\textbf
  {\bibinfo {volume} {17}},\ \bibinfo {pages} {L975} (\bibinfo {year}
  {1984})}\BibitemShut {NoStop}%
\bibitem [{\citenamefont {Weitz}\ and\ \citenamefont
  {Oliveria}(1984)}]{weitz:1984a}%
  \BibitemOpen
  \bibfield  {author} {\bibinfo {author} {\bibfnamefont {D.~A.}\ \bibnamefont
  {Weitz}}\ and\ \bibinfo {author} {\bibfnamefont {M.}~\bibnamefont
  {Oliveria}},\ }\href@noop {} {\bibfield  {journal} {\bibinfo  {journal}
  {Phys. Rev. Lett.}\ }\textbf {\bibinfo {volume} {52}},\ \bibinfo {pages}
  {1433} (\bibinfo {year} {1984})}\BibitemShut {NoStop}%
\bibitem [{\citenamefont {Weitz}\ \emph {et~al.}(1984)\citenamefont {Weitz},
  \citenamefont {Huang}, \citenamefont {Lin},\ and\ \citenamefont
  {Sung}}]{weitz:1984b}%
  \BibitemOpen
  \bibfield  {author} {\bibinfo {author} {\bibfnamefont {D.~A.}\ \bibnamefont
  {Weitz}}, \bibinfo {author} {\bibfnamefont {J.~S.}\ \bibnamefont {Huang}},
  \bibinfo {author} {\bibfnamefont {M.~Y.}\ \bibnamefont {Lin}}, \ and\
  \bibinfo {author} {\bibfnamefont {J.}~\bibnamefont {Sung}},\ }\href@noop {}
  {\bibfield  {journal} {\bibinfo  {journal} {Phys. Rev. Lett.}\ }\textbf
  {\bibinfo {volume} {53}},\ \bibinfo {pages} {1657} (\bibinfo {year}
  {1984})}\BibitemShut {NoStop}%
\bibitem [{\citenamefont {Chakrabarty}\ \emph {et~al.}(2009)\citenamefont
  {Chakrabarty}, \citenamefont {Moosm{\"u}ller}, \citenamefont {Arnott},
  \citenamefont {Garro}, \citenamefont {Tian}, \citenamefont {Slowik},
  \citenamefont {Cross}, \citenamefont {Han}, \citenamefont {Davidovits},
  \citenamefont {Onasch},\ and\ \citenamefont {Worsnop}}]{chakrabarty:2009}%
  \BibitemOpen
  \bibfield  {author} {\bibinfo {author} {\bibfnamefont {R.~K.}\ \bibnamefont
  {Chakrabarty}}, \bibinfo {author} {\bibfnamefont {H.}~\bibnamefont
  {Moosm{\"u}ller}}, \bibinfo {author} {\bibfnamefont {W.~P.}\ \bibnamefont
  {Arnott}}, \bibinfo {author} {\bibfnamefont {M.~A.}\ \bibnamefont {Garro}},
  \bibinfo {author} {\bibfnamefont {G.}~\bibnamefont {Tian}}, \bibinfo {author}
  {\bibfnamefont {J.~G.}\ \bibnamefont {Slowik}}, \bibinfo {author}
  {\bibfnamefont {E.~S.}\ \bibnamefont {Cross}}, \bibinfo {author}
  {\bibfnamefont {J.}~\bibnamefont {Han}}, \bibinfo {author} {\bibfnamefont
  {P.}~\bibnamefont {Davidovits}}, \bibinfo {author} {\bibfnamefont {T.~B.}\
  \bibnamefont {Onasch}}, \ and\ \bibinfo {author} {\bibfnamefont {D.~R.}\
  \bibnamefont {Worsnop}},\ }\href@noop {} {\bibfield  {journal} {\bibinfo
  {journal} {Phys. Rev. Lett.}\ }\textbf {\bibinfo {volume} {102}},\ \bibinfo
  {pages} {235504} (\bibinfo {year} {2009})}\BibitemShut {NoStop}%
\bibitem [{\citenamefont {Sander}\ \emph {et~al.}(2010)\citenamefont {Sander},
  \citenamefont {Patterson}, \citenamefont {Raj},\ and\ \citenamefont
  {Kraft}}]{sanderchakrabarty:2010}%
  \BibitemOpen
  \bibfield  {author} {\bibinfo {author} {\bibfnamefont {M.}~\bibnamefont
  {Sander}}, \bibinfo {author} {\bibfnamefont {R.~I.~A.}\ \bibnamefont
  {Patterson}}, \bibinfo {author} {\bibfnamefont {A.}~\bibnamefont {Raj}}, \
  and\ \bibinfo {author} {\bibfnamefont {M.}~\bibnamefont {Kraft}},\
  }\href@noop {} {\bibfield  {journal} {\bibinfo  {journal} {Phys. Rev. Lett.}\
  }\textbf {\bibinfo {volume} {104}},\ \bibinfo {pages} {119601} (\bibinfo
  {year} {2010})}\BibitemShut {NoStop}%
\bibitem [{\citenamefont {Chakrabarty}\ \emph {et~al.}(2010)\citenamefont
  {Chakrabarty}, \citenamefont {Moosm{\"u}ller}, \citenamefont {Arnott},
  \citenamefont {Garro}, \citenamefont {Tian}, \citenamefont {Slowik},
  \citenamefont {Cross}, \citenamefont {Han}, \citenamefont {Davidovits},
  \citenamefont {Onasch},\ and\ \citenamefont
  {Worsnop}}]{chakrabartyReply:2010}%
  \BibitemOpen
  \bibfield  {author} {\bibinfo {author} {\bibfnamefont {R.~K.}\ \bibnamefont
  {Chakrabarty}}, \bibinfo {author} {\bibfnamefont {H.}~\bibnamefont
  {Moosm{\"u}ller}}, \bibinfo {author} {\bibfnamefont {W.~P.}\ \bibnamefont
  {Arnott}}, \bibinfo {author} {\bibfnamefont {M.~A.}\ \bibnamefont {Garro}},
  \bibinfo {author} {\bibfnamefont {G.}~\bibnamefont {Tian}}, \bibinfo {author}
  {\bibfnamefont {J.~G.}\ \bibnamefont {Slowik}}, \bibinfo {author}
  {\bibfnamefont {E.~S.}\ \bibnamefont {Cross}}, \bibinfo {author}
  {\bibfnamefont {J.}~\bibnamefont {Han}}, \bibinfo {author} {\bibfnamefont
  {P.}~\bibnamefont {Davidovits}}, \bibinfo {author} {\bibfnamefont {T.~B.}\
  \bibnamefont {Onasch}}, \ and\ \bibinfo {author} {\bibfnamefont {D.~R.}\
  \bibnamefont {Worsnop}},\ }\href@noop {} {\bibfield  {journal} {\bibinfo
  {journal} {Phys. Rev. Lett.}\ }\textbf {\bibinfo {volume} {104}},\ \bibinfo
  {pages} {119602} (\bibinfo {year} {2010})}\BibitemShut {NoStop}%
\bibitem [{\citenamefont {Heinson}\ \emph {et~al.}(2010)\citenamefont
  {Heinson}, \citenamefont {Sorensen},\ and\ \citenamefont
  {Chakrabarti}}]{heinson:2010}%
  \BibitemOpen
  \bibfield  {author} {\bibinfo {author} {\bibfnamefont {W.~R.}\ \bibnamefont
  {Heinson}}, \bibinfo {author} {\bibfnamefont {C.~M.}\ \bibnamefont
  {Sorensen}}, \ and\ \bibinfo {author} {\bibfnamefont {A.}~\bibnamefont
  {Chakrabarti}},\ }\href@noop {} {\bibfield  {journal} {\bibinfo  {journal}
  {Aerosol Sci. Technol.}\ }\textbf {\bibinfo {volume} {44}},\ \bibinfo {pages}
  {i} (\bibinfo {year} {2010})}\BibitemShut {NoStop}%
\bibitem [{\citenamefont {Chakrabarty}\ \emph {et~al.}(2011)\citenamefont
  {Chakrabarty}, \citenamefont {Garro}, \citenamefont {Garro}, \citenamefont
  {Chancellor}, \citenamefont {Moosm{\"u}ller},\ and\ \citenamefont
  {Herald}}]{chakrabarty:2011a}%
  \BibitemOpen
  \bibfield  {author} {\bibinfo {author} {\bibfnamefont {R.~K.}\ \bibnamefont
  {Chakrabarty}}, \bibinfo {author} {\bibfnamefont {M.~A.}\ \bibnamefont
  {Garro}}, \bibinfo {author} {\bibfnamefont {B.~A.}\ \bibnamefont {Garro}},
  \bibinfo {author} {\bibfnamefont {S.}~\bibnamefont {Chancellor}}, \bibinfo
  {author} {\bibfnamefont {H.}~\bibnamefont {Moosm{\"u}ller}}, \ and\ \bibinfo
  {author} {\bibfnamefont {C.~M.}\ \bibnamefont {Herald}},\ }\href@noop {}
  {\bibfield  {journal} {\bibinfo  {journal} {Aerosol Sci. Technol.}\ }\textbf
  {\bibinfo {volume} {45}},\ \bibinfo {pages} {75} (\bibinfo {year}
  {2011})}\BibitemShut {NoStop}%
\bibitem [{\citenamefont {Martos}\ \emph {et~al.}(2017)\citenamefont {Martos},
  \citenamefont {Lapuerta}, \citenamefont {Exp{\'o}sito},\ and\ \citenamefont
  {{Sanmiguel-Rojas}}}]{martos:2017}%
  \BibitemOpen
  \bibfield  {author} {\bibinfo {author} {\bibfnamefont {F.~J.}\ \bibnamefont
  {Martos}}, \bibinfo {author} {\bibfnamefont {M.}~\bibnamefont {Lapuerta}},
  \bibinfo {author} {\bibfnamefont {J.~J.}\ \bibnamefont {Exp{\'o}sito}}, \
  and\ \bibinfo {author} {\bibfnamefont {E.}~\bibnamefont
  {{Sanmiguel-Rojas}}},\ }\href@noop {} {\bibfield  {journal} {\bibinfo
  {journal} {Powder Technol.}\ }\textbf {\bibinfo {volume} {311}},\ \bibinfo
  {pages} {528} (\bibinfo {year} {2017})}\BibitemShut {NoStop}%
\bibitem [{\citenamefont {Meakin}\ \emph {et~al.}(1985)\citenamefont {Meakin},
  \citenamefont {Vicsek},\ and\ \citenamefont {Family}}]{meakin:1985}%
  \BibitemOpen
  \bibfield  {author} {\bibinfo {author} {\bibfnamefont {P.}~\bibnamefont
  {Meakin}}, \bibinfo {author} {\bibfnamefont {T.}~\bibnamefont {Vicsek}}, \
  and\ \bibinfo {author} {\bibfnamefont {F.}~\bibnamefont {Family}},\
  }\href@noop {} {\bibfield  {journal} {\bibinfo  {journal} {Phys. Rev. B}\
  }\textbf {\bibinfo {volume} {31}},\ \bibinfo {pages} {564} (\bibinfo {year}
  {1985})}\BibitemShut {NoStop}%
\bibitem [{\citenamefont {Hasmy}\ \emph {et~al.}(1995)\citenamefont {Hasmy},
  \citenamefont {Foret}, \citenamefont {Anglaret}, \citenamefont {Pelous},
  \citenamefont {Vacher},\ and\ \citenamefont {Jullien}}]{hasmy:1995b}%
  \BibitemOpen
  \bibfield  {author} {\bibinfo {author} {\bibfnamefont {A.}~\bibnamefont
  {Hasmy}}, \bibinfo {author} {\bibfnamefont {M.}~\bibnamefont {Foret}},
  \bibinfo {author} {\bibfnamefont {E.}~\bibnamefont {Anglaret}}, \bibinfo
  {author} {\bibfnamefont {J.}~\bibnamefont {Pelous}}, \bibinfo {author}
  {\bibfnamefont {R.}~\bibnamefont {Vacher}}, \ and\ \bibinfo {author}
  {\bibfnamefont {R.}~\bibnamefont {Jullien}},\ }\href@noop {} {\bibfield
  {journal} {\bibinfo  {journal} {J. Non-Cryst. Sol.}\ }\textbf {\bibinfo
  {volume} {186}},\ \bibinfo {pages} {118} (\bibinfo {year}
  {1995})}\BibitemShut {NoStop}%
\bibitem [{\citenamefont {Lattuada}\ \emph
  {et~al.}(2003{\natexlab{a}})\citenamefont {Lattuada}, \citenamefont {Wu},\
  and\ \citenamefont {Morbidelli}}]{lattuada:2003a}%
  \BibitemOpen
  \bibfield  {author} {\bibinfo {author} {\bibfnamefont {M.}~\bibnamefont
  {Lattuada}}, \bibinfo {author} {\bibfnamefont {H.}~\bibnamefont {Wu}}, \ and\
  \bibinfo {author} {\bibfnamefont {M.}~\bibnamefont {Morbidelli}},\
  }\href@noop {} {\bibfield  {journal} {\bibinfo  {journal} {J. Coll. Inter.
  Sci.}\ }\textbf {\bibinfo {volume} {268}},\ \bibinfo {pages} {106} (\bibinfo
  {year} {2003}{\natexlab{a}})}\BibitemShut {NoStop}%
\bibitem [{\citenamefont {Lattuada}\ \emph
  {et~al.}(2003{\natexlab{b}})\citenamefont {Lattuada}, \citenamefont {Wu},
  \citenamefont {Hasmy},\ and\ \citenamefont {Morbidelli}}]{lattuada:2003}%
  \BibitemOpen
  \bibfield  {author} {\bibinfo {author} {\bibfnamefont {M.}~\bibnamefont
  {Lattuada}}, \bibinfo {author} {\bibfnamefont {H.}~\bibnamefont {Wu}},
  \bibinfo {author} {\bibfnamefont {A.}~\bibnamefont {Hasmy}}, \ and\ \bibinfo
  {author} {\bibfnamefont {M.}~\bibnamefont {Morbidelli}},\ }\href@noop {}
  {\bibfield  {journal} {\bibinfo  {journal} {Langmuir}\ }\textbf {\bibinfo
  {volume} {19}},\ \bibinfo {pages} {6312} (\bibinfo {year}
  {2003}{\natexlab{b}})}\BibitemShut {NoStop}%
\bibitem [{\citenamefont {Rottereau}\ \emph
  {et~al.}(2004{\natexlab{a}})\citenamefont {Rottereau}, \citenamefont {Gimel},
  \citenamefont {Nicolai},\ and\ \citenamefont {Durand}}]{rottereau:2004a}%
  \BibitemOpen
  \bibfield  {author} {\bibinfo {author} {\bibfnamefont {M.}~\bibnamefont
  {Rottereau}}, \bibinfo {author} {\bibfnamefont {J.~C.}\ \bibnamefont
  {Gimel}}, \bibinfo {author} {\bibfnamefont {T.}~\bibnamefont {Nicolai}}, \
  and\ \bibinfo {author} {\bibfnamefont {D.}~\bibnamefont {Durand}},\
  }\href@noop {} {\bibfield  {journal} {\bibinfo  {journal} {Eur. Phys. J. E}\
  }\textbf {\bibinfo {volume} {15}},\ \bibinfo {pages} {133} (\bibinfo {year}
  {2004}{\natexlab{a}})}\BibitemShut {NoStop}%
\bibitem [{\citenamefont {Rottereau}\ \emph
  {et~al.}(2004{\natexlab{b}})\citenamefont {Rottereau}, \citenamefont {Gimel},
  \citenamefont {Nicolai},\ and\ \citenamefont {Durand}}]{rottereau:2004b}%
  \BibitemOpen
  \bibfield  {author} {\bibinfo {author} {\bibfnamefont {M.}~\bibnamefont
  {Rottereau}}, \bibinfo {author} {\bibfnamefont {J.~C.}\ \bibnamefont
  {Gimel}}, \bibinfo {author} {\bibfnamefont {T.}~\bibnamefont {Nicolai}}, \
  and\ \bibinfo {author} {\bibfnamefont {D.}~\bibnamefont {Durand}},\
  }\href@noop {} {\bibfield  {journal} {\bibinfo  {journal} {Eur. Phys. J. E}\
  }\textbf {\bibinfo {volume} {15}},\ \bibinfo {pages} {141} (\bibinfo {year}
  {2004}{\natexlab{b}})}\BibitemShut {NoStop}%
\bibitem [{\citenamefont {{D{\'i}ez Orrite}}\ \emph {et~al.}(2005)\citenamefont
  {{D{\'i}ez Orrite}}, \citenamefont {Stoll},\ and\ \citenamefont
  {Schurtenberger}}]{diezOrrite:2005}%
  \BibitemOpen
  \bibfield  {author} {\bibinfo {author} {\bibfnamefont {S.}~\bibnamefont
  {{D{\'i}ez Orrite}}}, \bibinfo {author} {\bibfnamefont {S.}~\bibnamefont
  {Stoll}}, \ and\ \bibinfo {author} {\bibfnamefont {P.}~\bibnamefont
  {Schurtenberger}},\ }\href@noop {} {\bibfield  {journal} {\bibinfo  {journal}
  {Soft Matter}\ }\textbf {\bibinfo {volume} {1}},\ \bibinfo {pages} {364}
  (\bibinfo {year} {2005})}\BibitemShut {NoStop}%
\bibitem [{\citenamefont {Babu}\ \emph {et~al.}(2008)\citenamefont {Babu},
  \citenamefont {Gimel},\ and\ \citenamefont {Nicolai}}]{babu:2008}%
  \BibitemOpen
  \bibfield  {author} {\bibinfo {author} {\bibfnamefont {S.}~\bibnamefont
  {Babu}}, \bibinfo {author} {\bibfnamefont {J.~C.}\ \bibnamefont {Gimel}}, \
  and\ \bibinfo {author} {\bibfnamefont {T.}~\bibnamefont {Nicolai}},\
  }\href@noop {} {\bibfield  {journal} {\bibinfo  {journal} {Eur. Phys. J. E}\
  }\textbf {\bibinfo {volume} {27}},\ \bibinfo {pages} {297} (\bibinfo {year}
  {2008})}\BibitemShut {NoStop}%
\bibitem [{\citenamefont {Lindsay}\ \emph
  {et~al.}(1989{\natexlab{a}})\citenamefont {Lindsay}, \citenamefont {Klein},
  \citenamefont {Weitz}, \citenamefont {Lin},\ and\ \citenamefont
  {Meakin}}]{lindsay:1988}%
  \BibitemOpen
  \bibfield  {author} {\bibinfo {author} {\bibfnamefont {H.~M.}\ \bibnamefont
  {Lindsay}}, \bibinfo {author} {\bibfnamefont {R.}~\bibnamefont {Klein}},
  \bibinfo {author} {\bibfnamefont {D.~A.}\ \bibnamefont {Weitz}}, \bibinfo
  {author} {\bibfnamefont {M.~Y.}\ \bibnamefont {Lin}}, \ and\ \bibinfo
  {author} {\bibfnamefont {P.}~\bibnamefont {Meakin}},\ }\href@noop {}
  {\bibfield  {journal} {\bibinfo  {journal} {Phys. Rev. A}\ }\textbf {\bibinfo
  {volume} {39}},\ \bibinfo {pages} {3112} (\bibinfo {year}
  {1989}{\natexlab{a}})}\BibitemShut {NoStop}%
\bibitem [{\citenamefont {Lindsay}\ \emph
  {et~al.}(1989{\natexlab{b}})\citenamefont {Lindsay}, \citenamefont {Klein},
  \citenamefont {Weitz}, \citenamefont {Lin},\ and\ \citenamefont
  {Meakin}}]{lindsay:1989}%
  \BibitemOpen
  \bibfield  {author} {\bibinfo {author} {\bibfnamefont {H.~M.}\ \bibnamefont
  {Lindsay}}, \bibinfo {author} {\bibfnamefont {R.}~\bibnamefont {Klein}},
  \bibinfo {author} {\bibfnamefont {D.~A.}\ \bibnamefont {Weitz}}, \bibinfo
  {author} {\bibfnamefont {M.~Y.}\ \bibnamefont {Lin}}, \ and\ \bibinfo
  {author} {\bibfnamefont {P.}~\bibnamefont {Meakin}},\ }\href@noop {}
  {\bibfield  {journal} {\bibinfo  {journal} {Phys. Rev. A}\ }\textbf {\bibinfo
  {volume} {39}},\ \bibinfo {pages} {3112} (\bibinfo {year}
  {1989}{\natexlab{b}})}\BibitemShut {NoStop}%
\bibitem [{\citenamefont {Davidchack}\ \emph {et~al.}(2015)\citenamefont
  {Davidchack}, \citenamefont {Ouldridge},\ and\ \citenamefont
  {Tretyakov}}]{davidchack:2015}%
  \BibitemOpen
  \bibfield  {author} {\bibinfo {author} {\bibfnamefont {R.~L.}\ \bibnamefont
  {Davidchack}}, \bibinfo {author} {\bibfnamefont {T.~E.}\ \bibnamefont
  {Ouldridge}}, \ and\ \bibinfo {author} {\bibfnamefont {M.~V.}\ \bibnamefont
  {Tretyakov}},\ }\href@noop {} {\bibfield  {journal} {\bibinfo  {journal} {J.
  Chem. Phys.}\ }\textbf {\bibinfo {volume} {142}},\ \bibinfo {pages} {144114}
  (\bibinfo {year} {2015})}\BibitemShut {NoStop}%
\bibitem [{\citenamefont {{Miller III}}\ \emph {et~al.}(2002)\citenamefont
  {{Miller III}}, \citenamefont {Eleftheriou}, \citenamefont {Pattnaik},
  \citenamefont {Ndirango}, \citenamefont {Newns},\ and\ \citenamefont
  {Martyna}}]{miller:2002}%
  \BibitemOpen
  \bibfield  {author} {\bibinfo {author} {\bibfnamefont {T.~F.}\ \bibnamefont
  {{Miller III}}}, \bibinfo {author} {\bibfnamefont {M.}~\bibnamefont
  {Eleftheriou}}, \bibinfo {author} {\bibfnamefont {P.}~\bibnamefont
  {Pattnaik}}, \bibinfo {author} {\bibfnamefont {A.}~\bibnamefont {Ndirango}},
  \bibinfo {author} {\bibfnamefont {D.}~\bibnamefont {Newns}}, \ and\ \bibinfo
  {author} {\bibfnamefont {G.~J.}\ \bibnamefont {Martyna}},\ }\href@noop {}
  {\bibfield  {journal} {\bibinfo  {journal} {J. Chem. Phys.}\ }\textbf
  {\bibinfo {volume} {116}},\ \bibinfo {pages} {8649} (\bibinfo {year}
  {2002})}\BibitemShut {NoStop}%
\bibitem [{\citenamefont {Theodorou}\ and\ \citenamefont
  {Suter}(1985)}]{theodorou:1985}%
  \BibitemOpen
  \bibfield  {author} {\bibinfo {author} {\bibfnamefont {D.~N.}\ \bibnamefont
  {Theodorou}}\ and\ \bibinfo {author} {\bibfnamefont {U.~W.}\ \bibnamefont
  {Suter}},\ }\href@noop {} {\bibfield  {journal} {\bibinfo  {journal}
  {Macromolecules}\ }\textbf {\bibinfo {volume} {18}},\ \bibinfo {pages} {1206}
  (\bibinfo {year} {1985})}\BibitemShut {NoStop}%
\bibitem [{\citenamefont {Vym{\v e}tal}\ and\ \citenamefont {Vondr{\'a}{\v
  s}ek}(2011)}]{vymetal:2011}%
  \BibitemOpen
  \bibfield  {author} {\bibinfo {author} {\bibfnamefont {J.}~\bibnamefont
  {Vym{\v e}tal}}\ and\ \bibinfo {author} {\bibfnamefont {J.}~\bibnamefont
  {Vondr{\'a}{\v s}ek}},\ }\href@noop {} {\bibfield  {journal} {\bibinfo
  {journal} {J. Phys. Chem. A}\ }\textbf {\bibinfo {volume} {115}},\ \bibinfo
  {pages} {11455} (\bibinfo {year} {2011})}\BibitemShut {NoStop}%
\bibitem [{\citenamefont {Carpineti}\ and\ \citenamefont
  {Giglio}(1992)}]{carpineti:1992}%
  \BibitemOpen
  \bibfield  {author} {\bibinfo {author} {\bibfnamefont {M.}~\bibnamefont
  {Carpineti}}\ and\ \bibinfo {author} {\bibfnamefont {M.}~\bibnamefont
  {Giglio}},\ }\href@noop {} {\bibfield  {journal} {\bibinfo  {journal} {Phys.
  Rev. Lett.}\ }\textbf {\bibinfo {volume} {68}},\ \bibinfo {pages} {3327}
  (\bibinfo {year} {1992})}\BibitemShut {NoStop}%
\bibitem [{\citenamefont {Bibette}\ \emph {et~al.}(1992)\citenamefont
  {Bibette}, \citenamefont {Mason}, \citenamefont {Gang},\ and\ \citenamefont
  {Weitz}}]{bibette:1992a}%
  \BibitemOpen
  \bibfield  {author} {\bibinfo {author} {\bibfnamefont {J.}~\bibnamefont
  {Bibette}}, \bibinfo {author} {\bibfnamefont {T.~G.}\ \bibnamefont {Mason}},
  \bibinfo {author} {\bibfnamefont {H.}~\bibnamefont {Gang}}, \ and\ \bibinfo
  {author} {\bibfnamefont {D.~A.}\ \bibnamefont {Weitz}},\ }\href@noop {}
  {\bibfield  {journal} {\bibinfo  {journal} {Phys. Rev. Lett.}\ }\textbf
  {\bibinfo {volume} {69}},\ \bibinfo {pages} {981} (\bibinfo {year}
  {1992})}\BibitemShut {NoStop}%
\bibitem [{\citenamefont {Kim}\ and\ \citenamefont {Berg}(2000)}]{kim:2000a}%
  \BibitemOpen
  \bibfield  {author} {\bibinfo {author} {\bibfnamefont {A.~Y.}\ \bibnamefont
  {Kim}}\ and\ \bibinfo {author} {\bibfnamefont {J.~C.}\ \bibnamefont {Berg}},\
  }\href@noop {} {\bibfield  {journal} {\bibinfo  {journal} {Langmuir}\
  }\textbf {\bibinfo {volume} {16}},\ \bibinfo {pages} {2101} (\bibinfo {year}
  {2000})}\BibitemShut {NoStop}%
\bibitem [{\citenamefont {Puertas}\ \emph {et~al.}(2001)\citenamefont
  {Puertas}, \citenamefont {{Fern{\'a}ndez-Barbero}},\ and\ \citenamefont {{de
  las Nieves}}}]{puertas:2001}%
  \BibitemOpen
  \bibfield  {author} {\bibinfo {author} {\bibfnamefont {A.~M.}\ \bibnamefont
  {Puertas}}, \bibinfo {author} {\bibfnamefont {A.}~\bibnamefont
  {{Fern{\'a}ndez-Barbero}}}, \ and\ \bibinfo {author} {\bibfnamefont {F.~J.}\
  \bibnamefont {{de las Nieves}}},\ }\href@noop {} {\bibfield  {journal}
  {\bibinfo  {journal} {J. Chem. Phys.}\ }\textbf {\bibinfo {volume} {115}},\
  \bibinfo {pages} {5662} (\bibinfo {year} {2001})}\BibitemShut {NoStop}%
\bibitem [{\citenamefont {Kim}\ \emph {et~al.}(2003)\citenamefont {Kim},
  \citenamefont {Hauch}, \citenamefont {Berg}, \citenamefont {Martin},\ and\
  \citenamefont {Anderson}}]{kim:2003}%
  \BibitemOpen
  \bibfield  {author} {\bibinfo {author} {\bibfnamefont {A.~Y.}\ \bibnamefont
  {Kim}}, \bibinfo {author} {\bibfnamefont {K.~D.}\ \bibnamefont {Hauch}},
  \bibinfo {author} {\bibfnamefont {J.~C.}\ \bibnamefont {Berg}}, \bibinfo
  {author} {\bibfnamefont {J.~E.}\ \bibnamefont {Martin}}, \ and\ \bibinfo
  {author} {\bibfnamefont {R.~A.}\ \bibnamefont {Anderson}},\ }\href@noop {}
  {\bibfield  {journal} {\bibinfo  {journal} {J. Coll. Inter. Sci.}\ }\textbf
  {\bibinfo {volume} {260}},\ \bibinfo {pages} {149} (\bibinfo {year}
  {2003})}\BibitemShut {NoStop}%
\end{thebibliography}
\end{document}